\documentclass[11pt]{article}

\usepackage{amsmath,amssymb,epsfig,cite,color,verbatim,array,bm,amsfonts,enumerate,enumitem}
\usepackage{graphics,float}
\usepackage[colorlinks=true,citecolor=DarkOrchid,urlcolor=black,linkcolor=Blue]{hyperref}
\hypersetup{ linktoc=page}
\usepackage{blindtext}
\usepackage{microtype}
\usepackage[latin9]{inputenc}
\usepackage{authblk}
\everymath{\displaystyle}
\usepackage{stackrel} 
\usepackage{float}

\usepackage[caption = false]{subfig}

\numberwithin{equation}{section}
\allowdisplaybreaks

\usepackage[dvipsnames]{xcolor}

\title{ Scalarization-like mechanism through spacetime anisotropic scaling symmetry }
 
\author[1]{Alfredo Herrera-Aguilar \footnote{aherrera@ifuap.buap.mx} }
\author[1]{Daniel F. Higuita-Borja \footnote{dhiguita@ifuap.buap.mx} }
\author[2]{Julio A. M\'{e}ndez-Zavaleta\footnote{julioamz@mpp.mpg.de}} 

\affil[1]{ {\small {\it Instituto de F\'isica, Benem\'erita Universidad Aut\'onoma de Puebla,  Apdo. Postal J-48, C.P. 72570. Puebla, Mexico} }}
\affil[2]{ {\small {\it Max-Planck-Institut f\"ur Physik (Werner-Heisenberg-Institut),
F\"ohringer Ring 6, 80805 Munich, Germany}}}

\begin{document}

\maketitle

\begin{abstract} 
We present a new family of exact black hole configurations, which is a solution to a generalized Einstein-Maxwell-Dilaton setup in arbitrary dimension. These solutions are asymptotically Lifshitz for any dynamical critical exponent $z\geq 1$. It turns out that the existence of a nontrivial scalar field is a direct consequence of breaking the spacetime isotropic scaling symmetry. This black hole family accepts various interesting limits that link it to well-known solutions in both the isotropic and anisotropic cases. We study the thermodynamics of these field configurations showing that the first law is satisfied and providing the corresponding Smarr formula, both of these relations account for an electric contribution. Furthermore, we show that for a  certain parameter region, the anisotropic field configuration with a nonzero scalar field is thermodynamically preferred. This observation, together with a direct verification of the so-called scalarization conditions, suggest that the emergence of the dilaton field is due to a mechanism similar to spontaneous scalarization.
\end{abstract}
 
\tableofcontents

\section{Introduction}
\label{sec:intro}

A remarkable characteristic of the black hole solutions in Einstein's gravity is the fact that, despite the complexity involved, each configuration is defined by a reduced set of parameters. For asymptotically flat spacetimes, Israel initiated in the pursue of the so-called uniqueness theorems five decades ago \cite{Israel:1967wq}. In short, they state that the most general stationary and axisymmetric electrovacuum black holes with a positive mass reduce to the Kerr-Newman class \cite{Mazur:1982db}. In the same line the emergence of no-hair conjectures, first hinted by Ruffini and Wheeler \cite{Ruffini:1971bza}, takes care of sorting out the type of matter fields allowed to exist around regular black holes. As a result, increasing the number of effectively independent parameters is restricted to very specific circumstances. 
In more detail, the classical no-hair theorems later formalized by Bekenstein \cite{Bekenstein:1971hc} provide arguments --- independent of the gravitational dynamics --- asserting that black holes are characterized only by conserved charges subject to Gauss-like laws such as the mass, angular momentum and electromagnetic charge. Coupling a scalar field (either massless or massive), a Proca field or a massive spin-2 field around a stationary black hole is shown to fail. Further refinement on the arguments \cite{Heusler:1996ft} and extensions to explore different asymptotes and matter self-interactions appeared over the years, see \cite{Lahiri:1992vg} for a few examples. However, the reach and efficacy of no hair theorems are heavily dependent on the assumptions under which forged. Therefore, alongside the derivation of such conjectures, many counter examples of black holes or compact objects supporting nongauge fields in their exterior have been reported, the so-called hairy configurations.

When a field not associated to a conservation law is regular everywhere in the exterior of the black hole, it is called a hair.  The first and most celebrated candidates to be hairy configurations  were found not much after the no-hair conjectures were introduced. In \cite{Bronnikov:1973fh}, a conformal coupling between a scalar field and the curvature  proved an effective mechanism to evade the conjecture's axioms.  Similarly, in \cite{Bizon:1990sr} and later in \cite{Volkov:1998cc}  a family of so-called colored back holes were reported. In these cases, the spherical solutions are supported by a $SU(2)$ Yang-Mills field. It was afterwards shown that both \emph{a priori} appealing solutions suffer from undesirable properties such as instabilities under linear perturbations, \cite{Bronnikov:1978mx} and \cite{Bizon:1991nt} respectively. Further, an ADM mass formula for the solitons of the Einstein-Yang-Mills system in terms of black hole horizon properties was obtained by using the isolated horizon formalism \cite{Nucamendi2000}. These models for colored black holes settled the basis for general strategies to overcome the no-hair impossibilities previously established.  This fact, together with new theoretical frameworks, gave rise to a very active search for genuinely hairy black holes. For a nice review on the recent progress, we refer the reader to \cite{Anabalon:2012dw}.

Closely related to the hairy scenario,  a more recent phenomenon has gained plenty of attention from the community: spontaneous scalarization. Originally discussed in the context of neutron stars \cite{Damour:1993hw,Harada:1997mr}, it was noted that a tensor-scalar configuration could become nontrivial for a  specific range of parameters --- the total mass in that particular case. Furthermore, the finite scalar configuration turned to be preferable, energetically speaking, as compared to the scalar-free case. A detailed analysis and review can be found in \cite{Salgado:1998sg}.  It was spotted in the original works \cite{Damour:1993hw} that spontaneous scalarization can leave the weak gravity limit untouched and be fitted to manifests itself only in strong gravity processes.  This means of course that the standard solar system tests would be blind to it. Actually, the machinery is ready to compare and probe the possible scalarized nature of black holes. For instance, forthcoming gravitational wave interferometry --- the LISA observatory in particular --- seems capable of shedding some light on the issue \cite{Berti:2015itd}. In a similar fashion, the new achievements in black hole shadow imaging can, in principle, provide evidence on possible scalar imprints
\cite{Cunha:2015yba}.

Spontaneously scalarized spacetimes are field configurations that are smoothly connected with a scalar-free solution to a given theory. A generic characteristic of such setups is that over the scalar-free background, the propagation of the scalar field would exhibit a tachyonic mode. The mechanisms driving spontaneous scalarization come from an effective nonminimal coupling between the scalar field and either additional (matter) fields, the geometry or both. Such a setup is well captured by the action,
\begin{align} \label{eq:S_scalar}
S[g_{\mu\nu},\phi, \Psi ]=\int{ d^{D}x \sqrt{-g} \left[\frac{R-2\lambda}{2\kappa}+\mathcal{L}_{\text{SF}}\left(g_{\mu\nu}, \phi, \nabla_{\mu}\phi\right) +h(\phi)\mathcal{I}\left(g_{\mu\nu}, \Psi, R^{\alpha}_{\beta\mu\nu} \right)  \right] },
\end{align}
defined over $D$ spacetime dimensions in the presence of a cosmological constant $\lambda$ and where $\kappa$ is the Einstein's gravitational constant. The function $\mathcal{I}$ can depend on a standard Lagrangian for the fields $\Psi$ and the curvature, and the arbitrary function $h$ measures the nonminimal coupling  between the scalar field and it. Notice that the dynamics of the scalar can be as generic as desired and it is given by the Lagrangian $\mathcal{L}_{\text{SF}}$. During the past two decades numerous examples fitting in \eqref{eq:S_scalar} have been reported.   To mention a few of the most notorious, when $\mathcal{I}$ is taken as the Gauss-Bonnet term, static spherical black holes have been shown to scalarize for various nonminimal coupling functions  \cite{Doneva:2017bvd} as well as rotating configuration\cite{Dima:2020yac}. More related to this work is the case when $\mathcal{I}$ is a function of the Maxwell term, thus exhibiting scalarization when the scalar field is nonminimally coupled to a gauge field \cite{Herdeiro:2018wub,Fernandes:2019rez,Brihaye:2019kvj}. 
As a particular application, in  \cite{WT} it was argued that spontaneous scalarization in compact objects (polytropic fluids) is accompanied by a spontaneous violation of the weak energy condition. However, in \cite{Salgado2004} it was further shown that the negativeness of such energy densities is not generic of scalar-tensor theories of gravity by using realistic models of dense matter. Further examples and a review on the topic can be found in \cite{Sotiriou:2011dz} and \cite{Herdeiro:2015waa} respectively.

Very recently, in \cite{Herdeiro:2019yjy} the idea of achieving spontaneous scalarization through breaking scale invariance was worked out. There,  quantum corrections to the Einstein-Maxwell theory are added in the form of a quartic term in the field strength, breaking thus the natural Weyl invariance of Maxwell's theory. In this work, we find that a similar effect --- yet not identical, as later made precise --- can be achieved by breaking the spacetime isotropy, that is, the scaling symmetry between the timelike and spatial directions. Backgrounds with this property are called anisotropic, and the most iconic example is the Lifshitz spacetime which we proceed to describe.

A good motivation to delve into anisotropic spaces is the cornerstone problem of the quantization of gravity. There are various problems arising in the UV regimes of gravitational theories. Among them, the dimensions carried by the coupling (Newton's) constant $[G_{D=4}] \sim L^{2}$ spoil its renormalizability \cite{tHooft:1974toh}. This can be corrected to the case of dimensionless coupling at the cost of introducing an scaling asymmetry   between time and space, as explored in Horava-like gravities \cite{Horava:2009uw}. The natural setup for it to take place is a Lifshitz spacetime. Originally introduced in \cite{Kachru:2008yh}, it is understood as a gravitational dual to non-Lorentz invariant quantum field theories. Therein, the deviation from spacetime isotropy is measured by the so-called critical dynamical exponent $z$. The metric of a Lifshitz spacetime takes the form
\begin{align}\label{eq:LifshitzET}
g_{\text{Lifshitz}}=-\left(\dfrac{r}{l}\right)^{2z}dt^2+\dfrac{dr^2}{\left(\dfrac{r}{l}\right)^2}+\left(\dfrac{r}{l}\right)^2\sum^{D-2}_{k=1}dx^{2}_{k}.
\end{align}
In these coordinates, it turns out to be invariant under the anisotropic scaling transformations $D_z,$
\begin{equation}
 t\mapsto\lambda^zt,\qquad\,\quad r\mapsto\dfrac{r}{\lambda},\qquad\,\quad x_k\mapsto\lambda x_k,
\end{equation}
as well as under space and time translations and spatial rotations:  
\begin{eqnarray}\label{eqs}
H:&  & \,\,\, t \longrightarrow \,\,\,t'=t+a;\nonumber \\
P^{i}:&  & x^{i}\longrightarrow x^{'i}=x^{i}+a^{i}; \\
L^{ij}:&  & x^{i}\longrightarrow x^{'i}=L^{i}_{j}x^{j}.\nonumber 
\end{eqnarray}
Lifshitz spacetimes are anisotropic generalizations of anti-de Sitter (AdS) that also have constant negative curvature and share many of the geometrical and physical properties.  For instance, the existence of a boundary and the corresponding confining character as well as a Breitenlohner-Freedman (BF) bound for the energy --- or mass spectrum --- of fields dynamically coupled to gravity. We shall consider the parameter $z\geq1$ to be continuous for the purposes of our work.  Inviting to us, the idea that  Lorentz  invariance is  the low energy effect of a more fundamental anisotropic theory has been entertained \cite{Chadha:1982qq}. In the same line of thought, we can consider that the anisotropic  parameter $z$ that encodes the measure of Lorentz symmetry violation, achieves the relativistic limit when $z=1$, recovering the isotropy between space and time scaling transformations.

Naturally, while breaking the Lorentzian symmetry, Lifshitz backgrounds are not expected solutions to standard gravitational theories.  Therefore, the occurrence of nontrivial gravitational phenomena such as black holes requires either a modification on the dynamics or an enhancement through additional matter fields. For example, the first asymptotically Lifshitz back holes were reported in \cite{Taylor:2008tg, AyonBeato:2009nh, AyonBeato:2010tm}, one in the context of Einstein-Maxwell-Dilaton theory and the latter in gravities with quadratic corrections.

In the present  manuscript, we show how the spacetime anisotropy can be supplemented by a more generic matter arrangement than the Maxwell-Dilaton case. By assuming a simple static Lifshitz \emph{ansatz}, we find a novel exact charged black hole configuration. A nontrivial scalar field configuration manifests itself as one deviates from the isotropic case; i.e., $z>1$. As we will  explore in detail, these solutions are thermodynamically favorable under the existence of the scalar field. In this sense, we argue on favor of a scalarization phenomenom akin to the spontaneous scalarization  reviewed above. 
\\
\\
\emph{Organization of the paper.} \, In Sec.~\ref{sec:Sec1} we introduce the theoretical framework describing the nonminimal coupling between the scalar and a Maxwell field. For it, we find a four-dimensional solution shown to accommodate an event horizon in Sec.~\ref{sec:horizon}. Later, in Sec.~\ref{sec:higherD} we extend the result to a higher dimensional black hole and prove in Sec.~\ref{sec:scalarization_con} that it fulfils the scalarization conditions. The conserved charges and corresponding thermodynamics are provided in Sec.~\ref{Sec:Thermodynamics}. With these results, we finally  discuss thermodynamical arguments favoring the scalarized (anisotropic) configuration in Sec.~\ref{sec:scalarization_T}. We conclude with some important remarks and a discussion in Sec.~\ref{sec:remarks}.

\section{Lifshitz black holes \label{sec:Sec1} }

The existence of scalarized solutions conceded by an adequate source term in the scalar field equation. This source originates from the nonminimal coupling between the scalar mode and either additional matter or the curvature, as depicted in \eqref{eq:S_scalar}. For our purposes, we will  track down the first possibility. In particular,	 we will introduce a vector field $A_\mu$ such that $\mathcal{I}$ is chosen as the Maxwell's kinetic term.  Being the nonminimal coupling function arbitrary, our starting point results in a generalization of the standard Einstein-Maxwell-Dilaton theory in four spacetime dimensions\footnote{Pioneer research on black hole solutions to Dilaton gravity is reported in \cite{Gibbons:1987ps,Garfinkle:1990qj}. Also relevant in this context is the Lifshitz solution in \cite{Taylor:2008tg,AyonBeato:2009nh, AyonBeato:2010tm}. }. Namely, the action of our framework reads
\begin{align}\label{eq:action4D}
S\left[g_{\mu\nu},A_{\mu},\phi \right]=\int{  d^4x \sqrt{-g}  \Big( \frac{ R-2\lambda}{2\kappa}-\frac{1}{2}\partial_{\mu}\phi\partial^{\mu}\phi -\frac{ h(\phi)}{4} F_{\mu\nu} F^{\mu\nu}  \Big)  },
\end{align}
 where $F_{\mu\nu}:=\nabla_{\mu}A_{\nu}-\nabla_{\nu}A_{\mu}$ is the field strength and we recall that $h(\phi)$ is a nonminimal coupling function. 
 The coupling does not carry derivatives of the scalar so that the number of degrees of freedom is trivially preserved. Following the standard notation, $R$ is the scalar curvature, $\lambda$ is a cosmological constant, and $\kappa$ is the Einstein's gravitational constant. Notice that in order to preserve an smooth decoupling limit between the scalar and the vector, the minimal requirement is that  $h(\phi_{0})=\text{const}$ for a trivial (ground) value $\phi_{0}$ of the scalar field. When $h(\phi)=e^{-\alpha \phi}$, with $\alpha$ and arbitrary constant, the Einstein-Maxwell-Dilaton theory is recovered.

The field equations derived from~\eqref{eq:action4D} through the action principle  read
 \begin{align} \label{eq:FEs}
  \mathcal{E}_{\mu\nu}:=R_{\mu\nu}-\frac{1}{2}R g_{\mu\nu}+\lambda g_{\mu\nu}-\kappa T_{\mu\nu}=0, \qquad  \,\,\nabla_{\nu}\left( hF^{\mu\nu} \right) =0, \qquad \,\,
  \Box \phi -\frac{1}{4} \frac{dh }{d \phi}F_{\mu\nu}F^{\mu\nu} =0,
 \end{align}
where the total energy-momentum tensor contains information from both matter fields and it is conformed by
\begin{align}
 T_{\mu\nu}=T^{\phi}_{\mu\nu}+h T^{A}_{\mu\nu}, \qquad \qquad  \nabla_{\mu} T^{\mu\nu}=0,
\end{align}
with $T^{\phi}_{\mu\nu}$ the free scalar field energy-momentum tensor and $T^{A}_{\mu\nu}$ the one for the Maxwell field, i.e,
\begin{align}
T^{\phi}_{\mu\nu}= \partial_{\mu}\phi\partial_{\nu}\phi-\frac{1}{2}(\partial_{\alpha}\phi\partial^{\alpha}\phi) g_{\mu\nu}, \qquad \quad \ T^{A}_{\mu\nu}=F_{\mu\alpha}{F_{\nu}}^{\alpha}-\frac{1}{4}F_{\alpha\beta}F^{\alpha\beta} g_{\mu\nu}. 
\end{align} 
 We point out that the first scrutiny of this system in the context of scalarization was in \cite{Herdeiro:2018wub}. There, it was first observed that the nonminimal coupling between the scalar field and the Maxwell invariant can trigger the spontaneous scalarization of the asymptotically flat Reissner-Nordstr\"om black hole.  In this section, we will follow a similar approach by exploring solutions with the difference that we will build upon an asymptotically Lifshitz background.

\subsection{The  $D=4$ black hole \label{sec:Sol4D} }

In contrast to the numerical solutions \cite{Herdeiro:2018wub,Brihaye:2019kvj}  known to the model \eqref{eq:action4D}, we are interested in the  pursuit of exact configurations. It will suffice to take the static \emph{ansatz,} 
\begin{align}
g=- \left( \frac{r}{l} \right) ^{2z} f(r)dt^2+  \dfrac{dr^2}{\left( \dfrac{r}{l} \right) ^{2} f(r) }+\left( \dfrac{r}{l} \right) ^{2}\left( dx^2+dy^2 \right).
\end{align}
Cast in this manner, the metric is ensured to be asymptotically Lifshitz as long as  $\lim_{r\rightarrow\infty} f(r)=1$. 
It is well known that, for the Dilatonic coupling, the theory accepts a solution of this type supported by a purely electrical vector field and a logarithmic scalar field \cite{Taylor:2008tg}.
Following these lines,  here we construct the most general  nonminimal coupling $h(\phi)$ and its geometry describing static Lifshitz black holes.  
We	 consider the fields to inherit the spacetime isometries such that they are functions of the $r$ coordinate only. The vector potential is taken to be of the purely electrical form; thus, the field \emph{ans\"atze} are
\begin{align}
\phi=\phi(r), \qquad \quad A=A(r)dt.
\end{align}
    
Plugging it into the field equations \eqref{eq:FEs} one has that, as a consequence of this \emph{ansatz}, the Maxwell equations accept a trivial first integral,    
\begin{align} \label{eq:solAt_0}
\partial_{r}\left( \sqrt{-g}\, hF^{rt} \right)=0\quad \Rightarrow\quad A'(r)=\frac{ Q  }{ h} \left (\frac{ l}{r} \right)^{3-z},
\end{align}
being $Q$ an arbitrary integration constant which, for a minimal coupling $h(r)=1$, takes the role of the electric charge. 
For the rest of the equations, we can take simplifying linear combinations in order to solve the truly independent ones. For instance, the combinations
\begin{align}
\mathcal{E}^t_{t}-\mathcal{E}^r_{r}=&\left(\frac{r}{l}\right)^2 f \left[ \left( \phi' \right)^2 -\frac{ 2(z-1)}{r^2} \right]=0, \label{eq:EtmEr} \\
\mathcal{E}^r_{r}+\mathcal{E}^x_{x}=& \frac{1}{2}\left( \frac{r}{l}\right)^2 \left[  f''+\frac{3z+5}{r}f'+\frac{2(z+2)(z+1)}{r^2}f+\frac{4 \lambda l^2}{r^2}  \right]=0, \label{eq:EtpEx}
\end{align}
decouple the system as a simple first order nonlinear equation for the scalar field and an inhomogeneous Euler equation for the metric function. The general solution is
\begin{align}\label{eq:sol_f_phi_0}
 f(r)=-\frac{2 \lambda l^2}{(z+2)(z+1)}+\frac{a}{ r^{z+2} }+\frac{b}{ r^{2(z+1)} },  \qquad\quad \phi(r)=\sqrt{ \frac{2(z-1)}{\kappa} }\text{ln} \left( \frac{r}{l} \right)+\phi_{0}.  
\end{align}
 where $a$, $b$, and $\phi_{0}$ are arbitrary integration constants all carrying length units, and $z\geq1$ to preserve the reality of the scalar field. 
 
The remaining field equations constitute an algebraic condition for the nonminimal coupling function with the solution,
 \begin{align} \label{eq:sol_h_0}
  h(\phi)= - \frac{2\kappa (z+1) Q^2 l^6 e^{-\sqrt{ \frac{2\kappa}{(z-1)} }(\phi-\phi_{0} )} }{ (z-1)\lambda l^2 -bl^{-2(z+1)}z(z+1) e^{ -\sqrt{ \frac{2\kappa}{(z-1)} }(z+1)(\phi-\phi_{0})  }}  .
 \end{align}
 Notice that when the scalar acquires a trivial value $\phi=\phi_{0}$, the coupling indeed reduces to a finite constant. More interestingly, for very small values of the constant $b$ satisfying $bl^{-2(z+1)} \rightarrow 0$,  the coupling becomes dilatonic $h\sim \text{exp}( -2 \sqrt{2\kappa/(z-1)}(\phi-\phi_0) )$ , and the whole configuration reduces exactly to the Taylor black hole \cite{Taylor:2008tg}.
Note that the metric --- as well as the curvature invariants --- are singular at $r=0$. With the coupling function determined, the Maxwell field can be explicitly integrated from Eq.~\eqref{eq:solAt_0}, 
\begin{align}\label{eq:sol_A_4d}
A(r)&=-\frac{ b  }{2\kappa Ql^{2z+3}}\left( \frac{l}{r} \right)^{z} +  \frac{ (z-1)}{4\kappa Q l^3}\left( \frac{r}{l} \right)^{z+2}.
\end{align}

 Altogether, the solutions \eqref{eq:sol_f_phi_0}-\eqref{eq:sol_A_4d}  satisfy, in the present form, the whole set of equations of motion.  However, one needs to ensure the proper asymptotic limit of the metric $\lim_{r\rightarrow\infty}g=g_{\text{Lifshitz}}$. This imposes a constraint on the cosmological constant
 \begin{align}
  \lambda=-\frac{(z+2)(z+1)}{2 l^2}. \qquad \, 
 \end{align}

 As it is customary for Lifshitz black holes, the metric function depends in powers on the dynamical critical exponent. Nonetheless, most of the reported examples involve a simple binomial in $r$ such that the horizon is straightforwardly tractable; see, for instance, \cite{Taylor:2008tg}. In our case, we have a  trinomial with powers of $r$ which are not even guaranteed to be integers or rational numbers. Therefore a careful determination of the horizons is needed.

\subsection{Existence of an event horizon \label{sec:horizon} }
\emph{ General case.}
The metric function $f$ as given by Eq.~\eqref{eq:sol_f_phi_0} and specialized to the asymptotic condition takes the form
\begin{align}\label{eq:sol_f_phi_1}
 f(r)=1+\frac{a}{ r^{z+2} }+\frac{b}{ r^{2(z+1)} }.  
\end{align}
The existence of a event horizon requires that $f$ has a simple zero for a fixed value of $r\neq 0$. As it stands, we might regard $f$ as parametrized by the critical exponent $z$ such that it needs to be fixed in order to obtain a well-defined polynomial problem $f(r;z)=0$. For an integer value of $z=n$ with $n \in \mathbb{N}$, 
$f$ becomes a polynomial with $2(n+1)$ as the highest power. In principle, this case would have an exact radical solution up to $n=1$, according to the Abel-Ruffini theorem.  In any other case, given a specific value of $z$, the existence of the roots of $f$ has to be considered individually, requiring in most of the cases a numerical approach. Yet, it is possible to characterize the conditions under which the horizons exist. In the Appendix, we  determine that a maximum of two real roots of $f$ can exist. Thus the existence of an event horizon and a Cauchy horizon is hinted. 

\emph{ Ensuring  the existence of an event horizon.} 
 There is an alternative to single out one root $r=r_{H}$ of the metric function for an arbitrary value of the critical dynamical exponent.  This is achieved at the cost of fixing one of the integration constants in terms of the other, such that, with the particular redefinition
\begin{align} \label{eq:Sol_a}
 a=-\left[ 1+b\, r_{H}^{-2(z+1)} \right] {r_{H}}^{z+2},
\end{align}
$f(r_{H})=0$ is trivially satisfied.  Considering this information, one can show indeed that the vanishing of
\begin{align}\label{eq:sol_f_phi_h}
 f(r)=1-\left( \frac{r_{H}}{ r }\right)^{z+2}\left[ 1+b\, r_{H}^{-2(z+1)} \right] +\frac{b}{ r^{2(z+1)} },  
\end{align}
defines a null hyperspace whose generator satisfies the Frobenius integrability conditions.  Accordingly, there is an event horizon with radius  $r_{H}$. We will refer to the solution with condition \eqref{eq:Sol_a} implemented as the black hole configuration.

\subsection{Scalar-free $z=1$ solution \label{sec:z=1} }

Inspecting the equation for the scalar field, Eq.~\eqref{eq:EtmEr}, one can already see that the $z=1$ asymptotically AdS will enforce  a trivial scalar field configuration $\phi=\phi_{0} $. 
This is also evident from the field solution where the only condition one can impose to turn it off is the isotropic limit $z=1$.
Nonetheless, the limit is smooth as opposed to what the form of the coupling \eqref{eq:sol_h_0} suggests, and the system is still solvable. Concretely, we find
\begin{align} \label{eq:Sol_z=1}
f(r)=-\frac{ \lambda l^2}{3}+\frac{a}{r^3}+ \frac{ 2\kappa l^6 Q^2}{r^4},  \qquad A(r)=\frac{Q l^2}{r},  \qquad h(\phi)=1, \qquad \phi(r)=\phi_{0}.
\end{align}
up to a redefinition of the integration constant $b$ appearing in the anisotropic solution. In this case, if the cosmological constant is set to $\lambda=-3/l^2$, the asymptotic spacetime will be AdS. The configuration \eqref{eq:Sol_z=1} describes a known charged black hole dubbed as the planar Reissner-Nordstr\"om-AdS black hole (pRN-AdS)~\cite{Chamblin:1999tk}. The name planar comes from the spatial foliation at the horizon coming from the $(t,r)$-constant spatial sector $ds^2=\left( r/l \right)^2(dx^2+dy^2)$. We recall that the usual Reissner-Nordstr\"om spacetime has a $D-2$-dimensional sphere instead. For further details on black holes with a flat foliation, we refer the reader to \cite{Qadir:2006aj, Horowitz:2012nnc}.
 
\section{Higher-dimensional generalization \label{sec:higherD} }
The action in Eq.~\eqref{eq:action4D} can be straightforwardly generalized to an arbitrary dimension,  
\begin{align}\label{eq:actionD}
S[g,A,\phi]=\int{  d^Dx \sqrt{-g}  \Big( \frac{ R-2\lambda}{2\kappa}-\frac{1}{2}\partial_{\mu}\phi\partial^{\mu}\phi -\frac{ h(\phi)}{4} F_{\mu\nu} F^{\mu\nu}  \Big)  }.
\end{align}
On its own merit,  higher dimensional gravities are worth attention.  Yet, a more interesting motivation is  their connection with the gauge/gravity duality connecting a bulk $D$ dimensional theory with a $D-1$ boundary conformal theory.  The canonical example is the celebrated \emph{AdS/CFT correspondence} \cite{Maldacena:1997re}, where the gravitational dual resides in a type IIB string theory over $AdS_5\times S^5$. From the Lifshitz anisotropic side, the breaking of the Lorentz invariance allows one to  relate through duality nonrelativistic quantum systems as those pertinent for condensed matter physics \cite{Brynjolfsson:2009ct}. Therefore, higher-dimensional Lifshitz gravities open a window that extends the applicability of the gauge/gravity duality to a wide plethora of quantum systems under the so-called gravity/condensed matter theory correspondence \cite{Hartnoll:2016}. From now on we will focus only in the gravitational side and expect to push ahead the understanding of the dual quantum field theory side.

\subsection{The $D\ge3$ black hole\label{sec:D>3} }
We might proceed to explore the arbitrary dimensional generalization of the problem discussed in Sec.~\ref{sec:Sol4D}. The appropriate Lifshitz \emph{ansatz}  for this case is given by
\begin{align}
g=-\left(\dfrac{r}{l}\right)^{2z}f(r) dt^2+\dfrac{dr^2}{\left(\dfrac{r}{l}\right)^2f(r)}+\left(\dfrac{r}{l}\right)^2 \sum^{D-2}_{k=1}  dx^{2}_{k} .
\label{gbh}
\end{align}
Again, the asymptotic condition $\lim_{r\rightarrow \infty}f(r)=1$ secures the desired asymptotic behavior $\lim_{r\rightarrow \infty} g= g_{\text{Lifshitz}}$. 

The role of the dimension $D$ as an additional parameter does not turn out to be of greater complication. The field equations can be decoupled in a similar fashion as elaborated around \eqref{eq:EtmEr}-\eqref{eq:EtpEx}, resulting in the solution,
\begin{align}\label{eq:Sol_D}
f(r)&=1+\frac{a}{r^{z+D-2}}+\frac{b}{r^{2( z+D-3)}},\qquad \qquad \qquad \qquad \,\,\,
\phi(r)=\sqrt{ \frac{(D-2)(z-1)}{\kappa} } \text{ln}\left( \frac{r}{l} \right)+\phi_{0},\nonumber\\
A(r)&=-\frac{ b(D-2) }{4\kappa Ql^{2(D+z)-5}}\left( \frac{l}{r} \right)^{z+D-4} +  \frac{ (z-1)}{4\kappa Q l^3}\left( \frac{r}{l} \right)^{z+D-2}, \quad    \lambda=-\frac{(z+D-3)(z+D-2)}{2l^2}.
\end{align} 
Notice that, despite depending on the dimension, the functional dependence of $f$ on  $r$ is always decaying for $z>1,\, D>3$. The case $z=1$, $D=3$ is peculiar and is treated separately in the next section. We might apply the same reasoning of Sec.~\ref{sec:horizon} to ensure the existence of an event horizon. Doing so leads to the metric function describing the black hole,
\begin{align}\label{eq:Sol_f_D}
f(r)&=1-\left(\frac{r_{H}}{r}\right)^{z+D-2}\left(1+br_{H}^{-2(z+D-3)}  \right)+\frac{b}{r^{2( z+D-3)}}.
\end{align}

Additionally, the unique nonminimal coupling function which supports this  configuration is
 \begin{align} \label{eq:sol_h_D}
  h(\phi)=\frac{ -4\kappa (z+D-3) Q^2 l^{2(D-1)}  e^{-2\sqrt{  \frac{\kappa(D-2)}{(z-1)} }(\phi-\phi_{0} )} }{ 2(z-1)\lambda l^2 -bl^{-2(z+D-3)}(D-2)(z+D-3)(z+D-4) e^{ -2\sqrt{ \frac{\kappa}{(z-1)(D-2)} }(z+D-3)(\phi-\phi_{0})  }} .
 \end{align}

For arbitrary dimension $D>3$, we have the associated scalar-free solution when the isotropy of spacetime is recovered. This is, for $z=1$ the scalar field is trivialized $\phi=\phi_{0}$. Under that circumstances, the configuration takes the form,
\begin{align}\label{eq:sol_f_phi_D}
 f(r)=1+\frac{a}{ r^{D-1} }+\frac{b}{ r^{2(D-2) } },  \qquad \lambda&=-\frac{(D-2)(D-1)}{2l^2},\qquad  A(r)=-\frac{b(D-2) }{ 4 \kappa Q l^D r^{(D-3)} },
\end{align}
where the nonminimal coupling function becomes a constant, 
  \begin{align}
h(\phi)=\frac{ 4\kappa Q^2 l^{2(D - 1)} }{b (D - 2) (D - 3) },
 \end{align}
 which normalizes the charge to the standard Maxwell case,
 \begin{align}
 A(r)=-\frac{ Q}{ l (D-3) r^{(D-3)} }
 \end{align}
 Just as in the four-dimensional case, after some redefinition of constants, the isotropic limit  corresponds to the pRN-AdS black hole \cite{Chamblin:1999tk}. 
 It is worth mentioning that an interesting research line consists in exploring the implications of the constructed Lifshitz black hole solution (\ref{gbh})-(\ref{eq:sol_h_D}) for its dual quantum field theory within the gravity/condensed matter theory correspondence.

\subsection{Degenerate case: $z=4-D$ \label{sec:degenerate} }
When the dynamical exponent takes the particular value $z=4-D$, the metric function $f$ \eqref{eq:sol_f_phi_D} exhibits a degeneracy in the powers of the coordinate $r$. 
This is a consequence of the Euler equation governing the dynamics of the metric function. Under such a situation, $f$ acquires a logarithmic branch of solution. 

Considering that we are interested in Lifshitz spaces with $z\geq1$, this degeneracy occurs only for $D=2,3$ dimensions. We exclude $D=2$ as it is well known GR is not appropriate to model gravity in a $1+1$ framework.  Thus, the degeneracy occurs exclusively for $D=3$, where the dynamical exponent is set to $z=1$.  As pointed out in Eq.~\eqref{eq:Sol_z=1}, this value has the effect of turning off the scalar field regardless of the dimension. Under these considerations, the solution is
\begin{align} \label{eq:sol_deg}
f(r)=1+\frac{a}{r^2}+ \frac{ b\,\text{ln}(r)}{r^2},\qquad \lambda=-\dfrac{1}{l^2}, \qquad h(\phi)=-\frac{ 4\kappa l^4 Q^2}{b}, \qquad A(r)=-\frac{b \,\text{ln}(r)}{4\kappa l^3 Q }+A_{0}. 
\end{align}
 After some redefinition of constants, one finds that Eq.~\eqref{eq:sol_deg} corresponds to the charged (nonrotating) BTZ solution that was presented originally in \cite{Banados:1992wn} and with ulterior precision in \cite{Martinez:1999qi} after avid discussions in the literature.  A further discussion on the charged and rotating  BTZ solution is available in \cite{Carlip:1995qv,Garcia:1999py}.

\subsection{A hint of scalarization \label{sec:scalarization_con} }

As we have adverted from the general solution of the nonminimal coupling function, \eqref{eq:sol_h_D}, both the action and the solution are continuously connected to the pRN-AdS. However,  the more contemporary understanding of scalarization, in contrast to the standard hairy case,  demands a more thorough analysis of the different limits between the scalarized and the scalar-free configurations.  Thus, the aforementioned condition is just the first among three clearly stated in previous works, e.g. \cite{Astefanesei:2020qxk}. We prove next that our novel solution satisfies all of them as a first criterion to deem it scalarized or not.

\emph{ Smooth limit.} The formal condition that ensures the connectedness between the scalar-free solution and the configuration with a nontrivial scalar requires that the field equations \eqref{eq:FEs} decouple from the scalar mode without trivializing the vector configuration.  When the scalar acquires  a trivial value $\phi=\phi_{0}$, the scalar equation has the following implication:
\begin{align} \label{eq:raw_smooth_con}
\left. \left[\Box \phi  -\frac{1}{4}\left(\frac{dh}{d\phi}\right)F_{\mu\nu} F^{\mu\nu} \right] \right|_{\phi=\phi_{0}} =0 \qquad \Rightarrow \qquad \left. \left( \frac{dh}{d\phi} \right) \right|_{\phi=\phi_{0}} F_{\mu\nu}F^{\mu\nu}\stackrel{!}{=}0,
\end{align} 
which, in order to preserve a generic vector value $F_{\mu\nu}F^{\mu\nu}\neq0$, needs $\left. \left( \frac{dh}{d\phi} \right) \right|_{\phi=\phi_{0}}=0$.
The nonminimal coupling function \eqref{eq:sol_h_0} allows for the implementation of this condition at the cost of restricting the parameter space. Namely, fulfilling the relation,
\begin{align} \label{eq:b_smooth}
(z+D-3)(z + D-4)bl^{-2(z+D-3)}  + 2\lambda l^2=0,
\end{align}
solves condition \eqref{eq:raw_smooth_con} as desired. The whole expression \eqref{eq:b_smooth} is presented dimensionless such that the dimensions of $b$ are evident.  We shall refer to \eqref{eq:b_smooth} as the \emph{smoothness condition} hereafter.\\

\emph{Tachyonic modes over the scalar-free solution. } Again, the object of interest is the scalar field equation. We consider its perturbations around the ground value  $\phi=\phi_{0}$. If we also use the above information, then the only contributions up to linear order in the Klein-Gordon equation are
\begin{align}\label{eq:KG}
\left( \Box-m_{\text{eff}}^2\right)\delta \phi=0,\qquad where \qquad m_{\text{eff}}^2:=-\frac{1}{4} \left. \left( \frac{d^{2}h}{d\phi^{2} }       F_{\mu\nu}F^{\mu\nu} \right) \right|_{\phi=\phi_{0} , \, g=g_{z=1}}.
\end{align}
In arbitrary dimension, the effective mass of the field perturbations turns out to be 
\begin{align} \label{eq:m_eff}
 m_{\text{eff}}^2=-\frac{(D-2)(D-1)}{2l^2} \left( \dfrac{l}{r} \right)^{2(D-2)},
\end{align}
 revealing the existence of tachyonic modes (particles with their negative mass square) in the isotropic (scalar-free) solution. It is well known that these tachyons can arise in spacetimes with negative curvature (like AdS and Lifshitz), generating an instability if their squared mass falls below a negative bound. Thus, the allowed range of negative squared mass values is obtained by computing the Breitenlohner-Freedman bound, guaranteeing the energy positivity of the system and, hence, its stability \cite{Keeler:2012mb, Breitenlohner:1982bm}.   Though the  effective mass displayed in \eqref{eq:m_eff} seems of a local character,  the D'alembertian comes with the same power of $r$ in \eqref{eq:KG}; thus, the inhomogeneity to the Klein-Gordon equation ends up being a constant.
 
\emph{Consequence of the smoothness condition}. As noted before, our configuration \eqref{eq:Sol_D} is continuously connected with the dilatonic black hole \cite{Taylor:2008tg} through $b\rightarrow 0$. 
In this limit, the condition \eqref{eq:b_smooth} makes evident that the latter configuration, unlike our solution, does not accept a smooth limit to the scalar-free case. 
A second consequence of the smoothness condition  is due to the fixed value that the parameter $b$ attains. Recall from  \eqref{eq:sol_h_0}  that the coupling function incorporates $b$ explicitly. Therefore,  for our configuration to be supported,  $h=h(\phi;z)$ gets parametrized in terms of the critical dynamical exponent.  What it means is that for  a given $z$, not only the background is fixed but also the coupling constants, and thus the theory are. This characteristic will play later an important role in the interpretation of the scalarization effect in the present context.

\section{Global charges and black hole Thermodynamics\label{Sec:Thermodynamics}}

The thermodynamics of black holes with different to flat asymptotics has been a major field of investigation.  For our concrete case of interests, a Lifshitz asymptote, the study of the thermodynamics was initiated in\cite{Liu:2014dva} and vastly developed later \cite{Gim:2014nba}. In these works, it is determined how the critical dynamical exponent plays a major role in the form of the conserved charges. As a consequence, the first law and other relevant thermodynamical properties get modified by the spacetime anisotropy.

We proceed to compute the quantities characterizing the horizon. The Hawking temperature, defined in terms of the surface gravity $\tilde{\kappa}$, is given by means of the relation
\begin{equation} \label{eq:Hawking_Temp}
T_H \equiv \frac{\tilde{\kappa}}{2 \pi}, \qquad \text{with} \qquad 
\tilde{\kappa}^{2}= -\frac{1}{2}  \left(\nabla_{\mu}\chi_{\nu}\right)\left(\nabla^{\mu} \chi^{\nu}\right),
\end{equation}
where $\chi=\chi^{\mu}\partial_{\mu}=\partial_{t}$ is the generator of the event horizon.
Accordingly, the Hawking temperature associated with the $D$-dimensional configuration \eqref{eq:Sol_D} reads
\begin{equation}\label{eq:Temp}
T_{H} = \frac{1}{4 \pi} \frac{(z+D-2) r_{H}^z-b(z+D-4)r_{H}^{-(z+2D-6)}}{\ell^{z+1}},
\end{equation}
whereas the Wald entropy  results proportional to the event horizon area  $\mathcal{A}$ and takes the form,
\begin{equation}\label{eq:Entropy}
\mathcal{S} \equiv \frac{\mathcal{A}}{4 G_{D}} = \dfrac{V_{D-2}}{4 G_{D}} \left(\dfrac{r_{H}}{\ell}\right)^{D-2},
\end{equation}
where $G_{D}=\kappa/8\pi$ is the $D$-dimensional gravitational constant and $V_{D-2}$ represents the Euclidean volume of the spatial sector.\\ 
The electric charge can be computed  through a Gaussian integral over a spatial hypersurface $\Sigma$ at asymptotic infinity
\begin{equation}\label{eq:ElectricCharge}
Q_{\text{e}}=\int_{\Sigma}d^{D-2}x\sqrt{|\gamma|}n^\mu u^\nu h(\phi)F_{\mu\nu}=QV_{D-2},
\end{equation}
where $\gamma$ is the induced metric on $\Sigma$,  with $u$ and $n$ its timelike and spacelike normal unit vectors 
\begin{equation}\label{eq:NormalVectors}
 u^\mu=\dfrac{1}{\sqrt{f(r)}}\left(\dfrac{l}{r}\right)^z, \qquad \quad n^\mu=\sqrt{f(r)}\,\dfrac{r}{l}.
\end{equation}
The electrostatic potential associated with the Maxwell field is defined as
\begin{equation}\label{eq:ElectricPotential}
\Phi_e :=A(r_H)=\dfrac{1}{4lQ\kappa}\left[(z-1)\left(\dfrac{r_H}{l}\right)^{z+D-2}-(D-2)\dfrac{b}{l^{2(z+D-3)}}\left(\dfrac{r_H}{l}\right)^{4-z-D}\right]. 
\end{equation}
Next, in order to compute the total energy carried by our black hole configuration we exploit the so-called generalized ADT quasilocal method, foremost introduced in \cite{Kim:2013zha}. This method has been implemented and proved to be well suited in the construction of conserved charges in higher order gravity theories (see, for instance, \cite{Ayon-Beato:2015jga, BravoGaete:2017dso, Ayon-Beato:2019kmz,Bravo-Gaete:2020ftn}), as well as Einstein-Maxwell-Dilaton theories \cite{Hyun:2015tia}. 
In this method, the conserved charge corresponding to a killing vector field $\xi$ is given by
\begin{equation} \label{eq:conserved_charge}
\mathcal{Q} (\xi) = \int d^{D-2} x_{\mu \nu} \left(\Delta K^{\mu \nu}(\xi) -2 \xi^{[\mu}\int_{0}^{1} ds \Theta^{\nu]} (\xi,s) \right),
\end{equation}
with $d^{D-2}x_{\mu \nu} := \frac{\epsilon_{\mu \nu \mu_{1} \mu_{2}\ldots \mu_{D-2} }}{2(D-2)!} dx^{\mu_{1}} \wedge\ldots \wedge dx^{\mu_{D-2}}$. Here $\epsilon_{\mu \nu \mu_{1} \mu_{2}\ldots \mu_{D-2}}$ corresponds to the totally antisymmetric Levi-Civita symbol in $D-2$ dimensions. On the other hand, $s$ stands for a parameter allowing an interpolation of the black hole configuration between the solution of interest ($s=1$) and the asymptotic one  ($s=0$). Furthermore, $\Delta K^{\mu \nu} (\xi)= K^{\mu \nu}_{s=1} (\xi)-K^{\mu \nu}_{s=0} (\xi)$ stands for the total difference of the Noether potentials between the two end points of the path, $s=[0,1]$.  Finally, $\Theta^{\nu}$ is the surface term usually disregarded during the variation of the action but that has an instrumental role in the construction of the off shell Noether current \cite{Kim:2013zha}. In general, it has a nontrivial contribution to the quasilocal global charges and it is particularly important in our calculations of the mass.  
The Noether potential and the surface terms associated to our model are found to be 
\begin{align} \label{eq:NoetherP}
K^{\mu \nu}(\xi) &= 2 \sqrt{-g} \left[\frac{\nabla^{[\mu}\xi^{\nu]}}{2 \kappa}-\frac{1}{2}\frac{\partial \mathcal{L}}{\partial (\partial_{\mu} A_{\nu})}  \xi^{\sigma}A_{\sigma}\right],\\ 
\label{eq:surfaceT}
\Theta^{\mu}(\delta g, \delta \phi , \delta A ) &=  2 \sqrt{-g} \left(  \frac{g^{\alpha[ \mu }\nabla^{\beta]} \delta g_{\alpha \beta}}{2 \kappa}  +\frac{1}{2} \frac{\partial \mathcal{L}}{\partial (\partial_{\mu} A_{\nu})} \delta A_{\nu} + \frac{1}{2} \frac{\partial \mathcal{L}}{\partial (\partial_{\mu} \phi)} \delta \phi \right).
\end{align}
To compute the mass according to \eqref{eq:conserved_charge}, the  timelike Killing vector takes the form $\eta=\eta^{\mu}\partial_{\mu}=\partial_{t}$. This expression, together with the nontrivial contributions coming from \eqref{eq:NoetherP} and \eqref{eq:surfaceT}, lead subsequently to 
\begin{equation}\label{eq:Mass}
\mathcal{M}= \frac{V_{D-2}}{2 \kappa \ell} (D-2)\left[1+b \,r_{H}^{-2(z+D-3)}\right]\left(\dfrac{r_{H}}{\ell}\right)^{z+D-2}.
\end{equation}
By making use of the expressions \eqref{eq:Mass}, \eqref{eq:Temp}, \eqref{eq:Entropy}, \eqref{eq:ElectricPotential}, and \eqref{eq:ElectricCharge}, it is easy to check that the first law of thermodynamics is satisfied 
\begin{equation}\label{eq:FLaw}
d \mathcal{M} = T_{H} d \mathcal{S}+\Phi_e dQ_e, 
\end{equation}
where there is no effective contribution from the electric work since $Q_e$ is independent of $r_H$.\footnote{The Gibbs free energy ($\mathcal{G}$) representation of the first law $d\mathcal{G}=Q_{e}d\Phi_{e}-SdT_{H}$, indeed has an explicit electrical contribution.} However, it will be relevant for the Smarr-like formula below. The following integral relation between the thermodynamical quantities holds
\begin{equation}\label{eq:Smarr_raw}
\mathcal{M}-\left(\dfrac{D-2}{z+D-2}\right)T\mathcal{S}= b\,\dfrac{(D-2)(z+D-3)}{(z+D-2)}\dfrac{V_{D-2}}{\kappa l^{z+D-1}}r_{H}^{-(z+D-4)},
\end{equation}
which reduces to the previously reported \cite{Taylor:2008tg,Hyun:2015tia} when the parameter $b$  goes to zero. The last expression can be translated to a Smarr-like formula,
\begin{equation}\label{eq:SmarrFormula}
\mathcal{M}=\dfrac{D-2}{D-z}\left( T_{H} S-4Q_{\text{e}}\Phi_{\text{e}} \right).
\end{equation}
One  must observe that there is no contribution of a scalar charge neither in the first law or the Smarr formula, contrary to similar reported results \cite{Astefanesei:2020qxk}.  This is consequence of the nondecaying character of our field, but a more subtle analysis is needed since we are not in an asymptotically flat case.

\begin{figure}[H]\centering
  \includegraphics[width=0.75\linewidth]{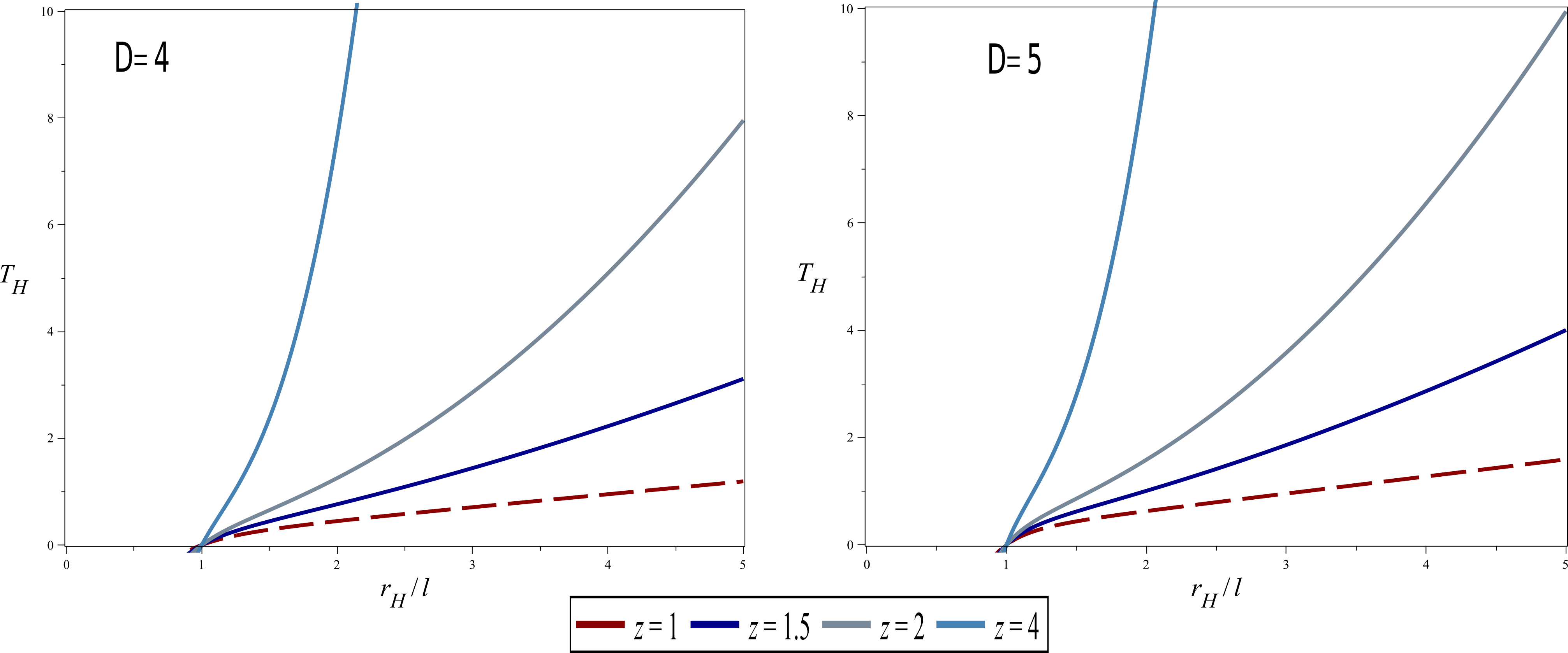} 
\caption{Behavior of the Hawking temperature ($T_{H}$) as a function of the normalized horizon radius ($r_{H}/l$). We plot two cases, $D=4$ (left) and $D=5$ (right), as well as different values of the critical exponent. We highlight the isotropic case $z=1$ with a dashed red line. Observe that, independent of the dimension, the temperature becomes positive for $r_{H}>l$ and is always increasing. This is true for all values of $z$.}
\label{fig:Tvsrh}
\end{figure}

Since the horizon radius is chosen as an independent and arbitrary parameter, there are no further reality or positivity restrictions  to impose. Thus, the charge $Q_{\text{e}}$, mass $\mathcal{M}$ or the coupling $b$ need not obey further restrictions. Nonetheless,  thermodynamical quantities still need to satisfy reasonable conditions such as a positive temperature and an always increasing entropy.  In terms of $r_{H}$, the later condition, the second law is well posed according to \eqref{eq:Entropy}. On the other hand, demanding a non-negative temperature puts a bound to the possible values of $r_{H}$, giving then a sense of domain of existence for physically meaningful black holes. From \eqref{eq:Temp}, we have that the condition,
\begin{align}\label{eq:domain_rh}
T_{H}\geq 0,\qquad \,\, \left( z+D-2 \right) \left[ \left( \frac{ r_{H}}{l}\right)^{2(z+D-3)}-1 \right]\geq 0,
\end{align}
implies that the domain of existence corresponds to $r_{H}\in [l, \infty) $. This region is depicted in Fig.~\ref{fig:Tvsrh}. Notice that the temperature is nondecreasing with the radius regardless of the dimension for $z\geq 1$.
 
\section{Lifshitz scalarization \label{sec:scalarization_T} }

In Sec.~\ref{sec:scalarization_con} we showed how our solution meets the sufficient conditions to be a scalarized version of the planar Reissner-Nordstrom-AdS black hole. This phenomenon can be further explored and made evident by the comparison of the thermodynamics of the scalar-free and hairy configurations.  For the following discussion,  recall that the	scalarized and scalar-free black holes are connected through the continuous limit $z\rightarrow 1$. 

We will consider the relevant thermodynamical quantities $\left( T, S, M \right) $ as given in \eqref{eq:Temp}, \eqref{eq:Entropy} and \eqref{eq:Mass} respectively.  For simplicity, we will take scale factor $l$ and the integration volume $V$ as being fixed to a unitary value. Bear in mind that the coupling constant $b$ has to satisfy the smoothness condition \eqref{eq:b_smooth}. The idea is to determine the effect of the dynamical critical exponent in the evolution of the thermodynamical variables; thus, we will leave $z$ as the parameter characterizing the thermodynamics.   
\begin{center}
\begin{figure}[h!]\centering
  \includegraphics[width=0.85\linewidth]{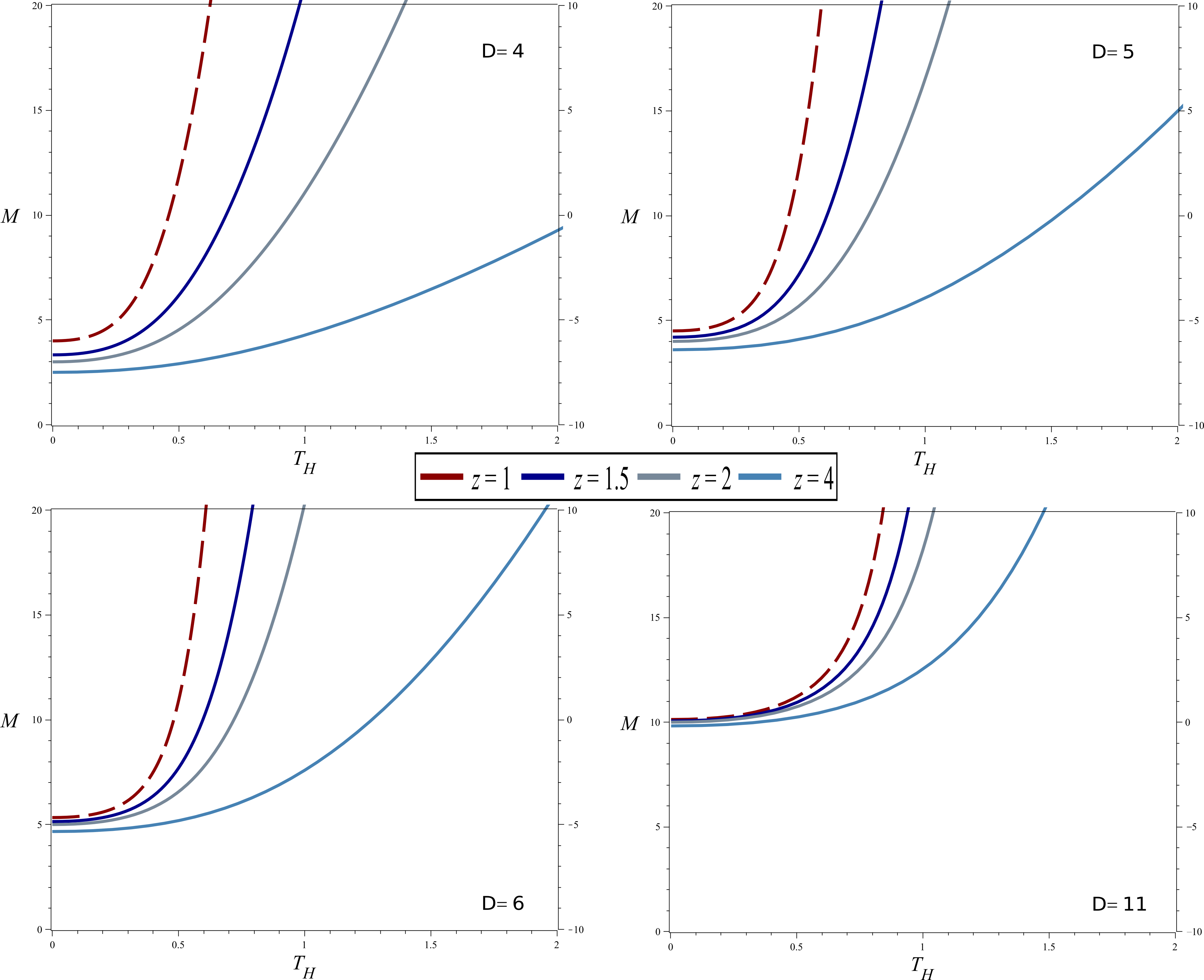}
  \caption{We display the energy as a function of the temperature taking all other thermodynamical quantities fixed. In order, from left to right, we display different dimensions: $D=4,5,6,11$. The dashed red line pictures the isotropic case $z=1$. The scalarized solutions appeared energetically favored in all the region $T_{H}\geq 0$ since they have a lower black hole mass.}
  \label{fig:MvsT}
\end{figure}
\end{center}
First, in Fig.~\ref{fig:MvsT} we display the curves of energy versus temperature for different values of $z$ and different dimensions. The most remarkable information provided by this plot is the fact that the energy for a given temperature value --- with the rest of the variables fixed ---  is always higher for the isotropic (scalar-free) black hole with respect to any other scalarized  ($z>1$) configuration.

\begin{align}
 \mathcal{M}(T; z=1) > \mathcal{M}(T;z>1).
\end{align}
An extension of the plot Fig.~\ref{fig:MvsT} to the region of negative temperatures displays a crossing point below which, the behavior is the inverse. Of course, we discard this region in compliance with the third law.

\begin{center}
\begin{figure}[h!]\centering
  \includegraphics[width=0.85\linewidth]{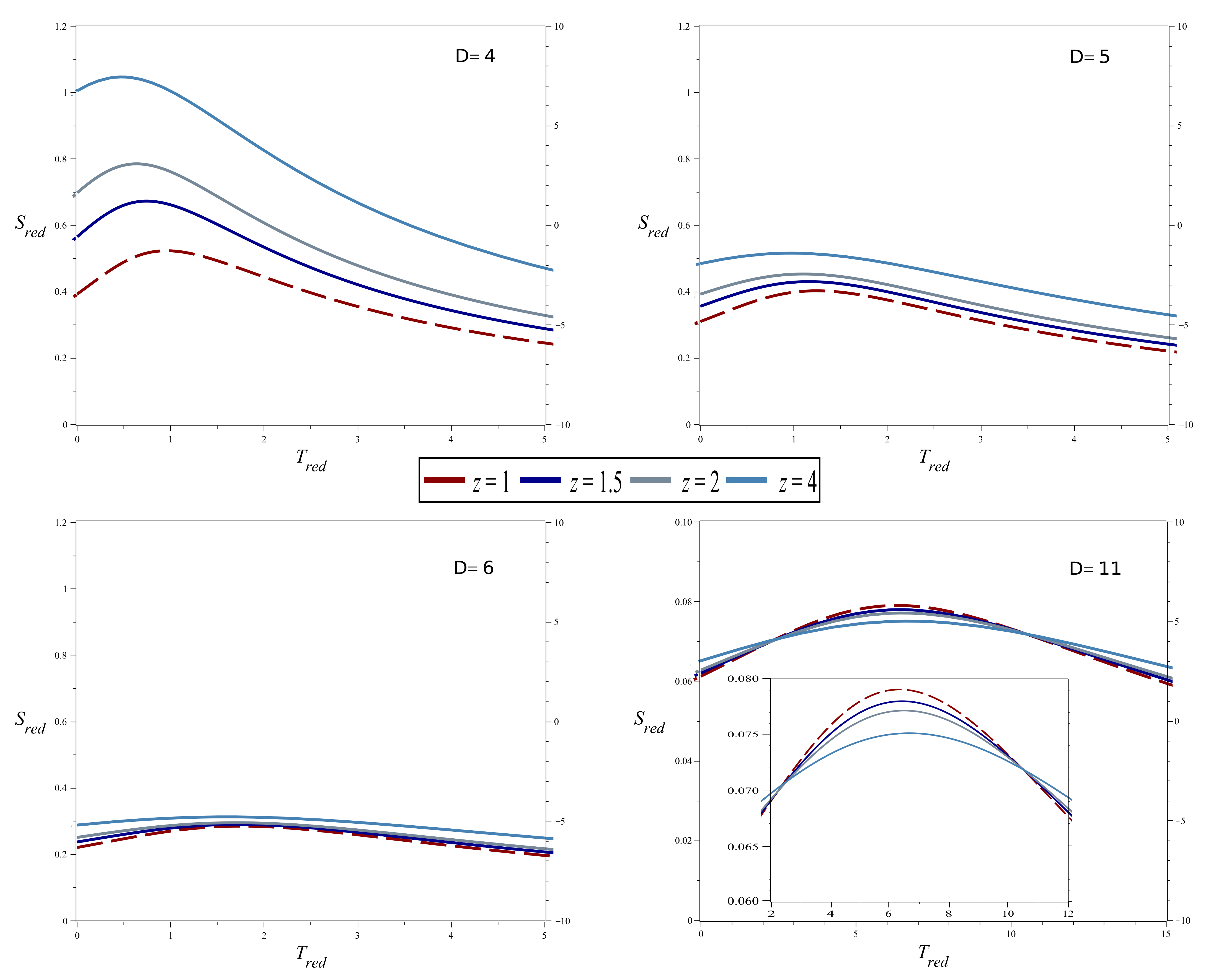}
  \caption{We plot the reduced entropy as a function of the reduced temperature while all other thermodynamical quantities are fixed. In order, from left to right, we display different dimensions: $D=4,5,6,11$. The dashed red line pictures the isotropic case $z=1$. In this picture,  scalarized solutions are entropically favored in all the region $T_{H}\geq 0$ for $D=4,5,6$. In $D=11$,  there is a small region of interchange which is shown zoomed in.}
  \label{fig:SrvsTr}
\end{figure}	
\end{center}

Looking for supplementary evidence, we portray the curves of reduced entropy versus reduced temperature in Fig.~\ref{fig:SrvsTr}, similarly as done in \cite{Fernandes:2019rez, Astefanesei:2020qxk}. These quantities are defined as
\begin{align}
T_{\text{red}}:=\mathcal{M}T_{H}, \qquad \quad S_{\text{red}}:=\frac{S}{\mathcal{M}^2}.
\end{align}
  Probing different dimensions, we realized that for  $D=4,5,6$ the reduced entropy at given value of the reduced temperature is always higher in the scalarized (anisotropic) configuration compared to the scalar-free. This result is in agreement with the qualitative behavior of the energy curves, indicating that the anisotropic configuration are preferred, 
\begin{align}
\mathcal{S_{\text{red}}}(T_{\text{red}}; z=1) < \mathcal{S_{\text{red}}}(T_{\text{red}};z>1). 
\end{align} 
  However, the plot in $D=11$ shows an exchange region where the behavior is interchanged but it is recovered soon after. The crossing places of all different curves correspond to a single point which suggest a physical transition that urges for a deeper analysis.   

The lower dimensional solution $D=3$ is left aside of the scalarization discussion. There is a main drawback  in that case originated by the possible degeneracy of the scalar-free solution as discussed in Sec.\ref{sec:degenerate}. Actually, it is reflected in the smoothness condition \eqref{eq:b_smooth}, which breaks down in said dimension and for $z=1$.  Therefore, the comparison of the thermodynamical quantities for isotropic and anisotrpic configurations might be not well posed.

\section{Conclusions and final remarks\label{sec:remarks} }

As a solution to a generalized Einstein-Maxwell-Dilaton theory, we have constructed a new family of exact Lifshitz black hole configurations. In such configurations, the existence of a nontrivial scalar field is a direct consequence of breaking the isotropic scaling symmetry between space and time ($z>1$). The scalar field trivializes at $z=1$ and becomes imaginary for $z<1$. The scalar-vector nonminimal coupling is mediated by the function  \eqref{eq:sol_h_D}, which is the most general for the given setup: a one-function static metric \emph{ansatz}, and a purely electric vector potential.
The obtained coupling can be understood as a nonlinear deviation from the standard  dilaton coupling measured by the order of the new parameter $b$. In contrast to the limiting solution \cite{Taylor:2008tg}, the generalized dilaton coupling upholds a second horizon in our black hole, allowing then extremal configurations. It turns out that this solution passes the minimal tests to exhibit (spontaneous) scalarization at the cost of fixing the value of $b$ according to the smoothness condition \eqref{eq:b_smooth}. If it is imposed, the dilaton limit is lost and the effective couplings of the action obtain an unavoidable dependence on the critical  exponent. 

We have shown how the anisotropic  black hole with a nontrivial scalar is energetically favored and, for a certain parameter region, it is also entropically favored as compared to the gravitating configuration with no scalar field.  The region where the configurations are physically meaningful follows from the imposition of positive temperature, which exclude undesired thermodynamical behavior such as ill-defined physical quantities. Both cases, the one with a nontrivial scalar and the scalar-free black holes are continuously connected through the limit $z\rightarrow1$. The other way around, the sudden emergence of a scalar field and a  preferred thermodynamics are driven by a deviation from the isotropic scaling symmetry measured by $z>1$. More precisely, the amplitude of the scalar field and the entropy grow with a larger critical exponent. This phenomenon is remarkably similar to that of \emph{spontaneous scalarization} where the evolution of the parameter into a threshold region triggers the transition of the configuration to a scalarized one. However the analogy is not exact.  To speak of spontaneous scalarization, the smoothness condition should be fulfilled, and therefore $z$ will be effectively introduced in the action. As a result, varying the critical exponent means also changing the theory, and the scalarization process would not take place in the same moduli space. 

Likewise, by regarding the dynamical exponent $z$ as a continuous parameter that can be varied towards the critical relativistic value $z=1$ --- in which the scalar field vanishes --- one can span several nonrelativistic quantum field theories with different physical properties. Since the scalarization phenomenon involves the emergence of a second order phase transition for the perturbations of the scalar field, it is interesting to explore its implications within the framework of the gravity/condensed matter theory correspondence. Work is currently in progress to elucidate the field theory physical meaning of this remarkable effect.

Despite the subtlety impeding the identification of the presented result as an analytical example of spontaneous  black hole scalarization, our result suggests that breaking the isotropic scaling can be also a source for this effect such as rotation \cite{Dima:2020yac} or the violation of conformal invariance \cite{Herdeiro:2019yjy}. Of course, there are many directions  to be explored as possible candidates capable of making compatible such scenarios. The most immediate one would be generalizing the scalar field dynamics, with the easiest case of including a self-interaction potential. This additional freedom could in principle remove the $z$ dependence from the action and allow the coexistence of both the scalarized and the scalar-free solution for a fixed coupling function. 

Some words are in order relating to the stability of the quite general family of scalarizing asymptotically Lifshitz black hole configurations here constructed. In this regard, in \cite{Giacomini:2012hg} the authors studied real scalar field fluctuations about a $D$-dimensional family of black holes with Lifshitz asymptotics and generic critical exponent $z$. They showed that the latter are stable under scalar field perturbations for suitable boundary conditions. 
The study was performed by assuming a stationary separable ansatz for a probe scalar field in the background of an asymptotically Lifshitz black hole. 
The strategy was to compute the quasinormal mode frequencies  and to impose incoming modes boundary conditions at the horizon and Dirichlet boundary conditions --- equivalent to a vanishing scalar field --- at infinity. 
It turns out that the imaginary part of these frequency modes always becomes negative, yielding a  damping-off scalar field perturbation, provided a suitable Breitenlohner-Freedman bound is satisfied. Thus the stability of Lifshitz black holes is secured under this kind of perturbations. 
A similar result was accomplished in \cite{Hashemi:2019jlt} for $D$-dimensional dilaton Lifshitz black holes coupled to Born-Infeld electrodynamics. Moreover, in \cite{CuadrosMelgar:2011up} the authors numerically probed the stability of a $D = z = 3$ Lifshitz black hole under both scalar and spinorial perturbations making use of quasinormal mode frequencies.
Our asymptotically Lifshitz black hole field configuration possesses the required properties to ensure the kind of stability proved in \cite{Giacomini:2012hg} since our scalar field effective mass also obeys a Breitenlohner-Freedman bound.
However, a complete stability analysis of  asymptotically Lifshitz black hole spacetimes requires addressing the tensor perturbations as well, an interesting open question that deserves further attention. In this sense, we would like to mention that it seems feasible to carry out such an enterprise given that our gravitational action is simply the Einstein-Hilbert term. 
The task would become more arduous  when considering  higher-curvature terms,  involving ghost fields, that have been considered previously in the literature.

\section*{Acknowledgements \centering}
All authors thank SNI for support. All authors are indebted to U. Nucamendi, E. Ay\'on-Beato, and M. Hassa\"ine for useful discussions. JAMZ thanks valuable comments from M. L\"uben. AHA and DFHB acknowledge financial support by a CONACYT Grant No. A1-S-38041. DFHB is also grateful to CONACYT for a {\it Postdoc por Mexico} Grant No. 372516. JAMZ is supported by a grant of the Max Planck Society.

\appendix

\section{General analysis of the horizons in $D=4$ \label{sec:appA}}

We recall the generic form of the metric function preserving the Lifshitz asymptote in four dimensions
\begin{align} \label{eq:f_gen_4}
f(r)=1+a \left( \dfrac{l}{r} \right)^{z+2}-b \left( \dfrac{l}{r} \right)^{2(z+1)},
\end{align}
where we have stripped the integration constant of length dimensions as compared to \eqref{eq:sol_f_phi_0} and we chose $b\rightarrow -b$ without loss of generality. As we have already mentioned, it is not possible to give closed form expressions for all the roots of $f$ due to the indefinite character of the powers of $r$. Actually, there are few known examples of asymptotically Lifshitz geometries that could admit more than one horizon. For instance, the solutions presented in \cite{Bravo-Gaete:2017nkp}  constitute another case.
Anyhow, studying the critical points \eqref{eq:f_gen_4} is still handleable. Taking the first derivative of  \eqref{eq:f_gen_4} yields two extrema, 
\begin{align}
r_{\text{ext}}=\left\{0,\,\, \left(-\dfrac{2a(z+1)}{b(z+2)}\right) ^{1/z} l \right\}.
\end{align}
Clearly, the first value is excluded because it coincides with the singularity.  We proceed to evaluate the second derivative of $f$ in the only sensible critical point 
\begin{align} \label{eq:min_con}
\left. \dfrac{d^2f}{dr^2}\right|_{r=r_{\text{ext}} } = \frac{1}{2}\left(\dfrac{b}{al}\right)^2\left(\dfrac{2a}{b} \right)^{-4/z} \dfrac{(z+2)^{2(z+2)/z} }{(z+1)^{(z+4)/z} } \left[ a^2(z+3)-b^2(2z+3)  \right].
\end{align}
Observe that to determine the sign the above expression, we can factor out one of the integration constants, say $a$ for example, such that the only relevant quantity is their ratio $\beta:=b/a$. In order to generate at least one horizon at $r_{H}>0$ , the only physically acceptable case is for $r_{\text{ext}}$ to be a minimum, demanding then 
\begin{align}
z+3-\beta^2(2z+3) \beta  >0 \quad \Rightarrow \quad -\sqrt{\frac{2+3}{2z+3}}<\beta<\sqrt{\frac{2+3}{2z+3}}.
\end{align}
Additionally, it is necessary that $f(r_{\text{ext}})\leq 0$ such that there is at least one real root.  When $f(r_{\text{ext}}) = 0$ there is  one unique event horizon given by $r_{H}=r_{ext}$.  The case allowing for two roots is met when $f(r_{\text{ext}})<0$.
If this condition and \eqref{eq:min_con}  are simultaneously satisfied the black hole displays both, an event horizon and a Cauchy horizon.


\end{document}